\documentclass[%
 aip,
 amsmath,amssymb,
 reprint,%
]{revtex4-1}

\usepackage[english]{babel}
\usepackage[utf8]{inputenc}
\usepackage{SIunits}
\usepackage{graphicx}
\usepackage{dcolumn}
\usepackage{amsmath}
\usepackage{natbib}
\usepackage{bm}
\usepackage{textcomp}
\usepackage{color}

\begin{document}

\title{Magneto-spectroscopy of exciton Rydberg states in a CVD grown WSe$_2$ monolayer}

\author{A. Delhomme}
\affiliation{LNCMI (CNRS, UJF, UPS, INSA), BP 166, 38042 Grenoble Cedex 9, France}

\author{G. Butseraen}
\affiliation{Univ. Grenoble Alpes, CNRS, Grenoble INP, Institut Néel, 38000 Grenoble, France}

\author{B. Zheng}
\affiliation{Key Laboratory for Micro-Nano Physics and Technology of Hunan Province, State Key Laboratory of Chemo/Biosensing and Chemometrics, College of Materials Science and Engineering, Hunan University, Changsha, Hunan 410082, China}

\author{L. Marty}
\affiliation{Univ. Grenoble Alpes, CNRS, Grenoble INP, Institut Néel, 38000 Grenoble, France}

\author{V. Bouchiat}
\affiliation{Univ. Grenoble Alpes, CNRS, Grenoble INP, Institut Néel, 38000 Grenoble, France}

\author{M.R. Molas}
\affiliation{Institute of Experimental Physics, Faculty of Physics, University of Warsaw, ul. Pasteura 5, 02-093 Warszawa, Poland}

\author{A. Pan}
\affiliation{Key Laboratory for Micro-Nano Physics and Technology of Hunan Province, State Key Laboratory of Chemo/Biosensing and Chemometrics, College of Materials Science and Engineering, Hunan University, Changsha, Hunan 410082, China}

\author{K. Watanabe}
\affiliation{National Institute for Materials Science, Tsukuba, 305-0044, Japan}

\author{T. Taniguchi}
\affiliation{National Institute for Materials Science, Tsukuba, 305-0044, Japan}

\author{A. Ouerghi}
\affiliation{Centre de Nanosciences et de Nanotechnologies, CNRS, Univ. Paris-Sud, Université Paris-Saclay, C2N – Marcoussis, 91460 Marcoussis, France }

\author{J. Renard}
\affiliation{Univ. Grenoble Alpes, CNRS, Grenoble INP, Institut Néel, 38000 Grenoble, France}

\author{C. Faugeras}
\email{clement.faugeras@lncmi.cnrs.fr} \affiliation{LNCMI (CNRS, UJF, UPS, INSA),
BP 166, 38042 Grenoble Cedex 9, France}

\date{\today }

\begin{abstract}
The results of magneto-optical spectroscopy investigations of excitons in a CVD grown monolayer of WSe$_2$ encapsulated in hexagonal boron nitride are presented. The emission linewidth for the $1$s state is of $4.7$~meV, close to the narrowest emissions observed in monolayers exfoliated from bulk material. The $2$s excitonic state is also observed at higher energies in the photoluminescence spectrum.  Magneto-optical spectroscopy allows for the determination of the g-factors and of the spatial extent of the excitonic wave functions associated with these emissions. Our work establishes CVD grown monolayers of transition metal dichalcogenides as a mature technology for optoelectronic applications.
\end{abstract}

\maketitle

Monolayers of semiconducting transition metal dichalcogenides (S-TMD), labelled as MX$_2$, where M=W, Mo, Re or Zr and X=S, Se or Te for the most studied compounds, are direct band gap semiconductors with the band gap located at the two inequivalent $K^{\pm}$ points of their hexagonal Brillouin zone~\cite{mak2010}. They exhibit a large number of properties interesting both for fundamental science and for technology~\cite{koperski}. These properties include a band gap in the visible range~\cite{Tonndorf2013}, a strong light matter coupling~\cite{mak2016}, the presence of single photon emitters~\cite{Koperski2015,He2015}, very strong excitonic effects~\cite{he, stier2018,molas2019}, possibilities of tuning the emission energy by Coulomb engineering~\cite{Stier2016, Raja2017}, and the opportunity of combining them to obtain type II band alignment~\cite{Chiu2015}. The spin-orbit interaction in these materials is strong and splits the spin levels in the valence band by few hundreds of meV, and by few tens of meV in the conduction band~\cite{Xu2014}. This large splitting in the valence band defines two different excitons at the K$^{\pm}$ points attached to the two spin-split bands, named A and B excitons~\cite{arora2015}. Because of the lack of inversion symmetry, the two different valleys can be excited independently using circularly polarized light~\cite{mak2012}, opening the possibility of initializing an exciton population in a given valley, or creating coherent superpositions of both valleys~\cite{jones}.

Best specimens of S-TMD monolayers are usually obtained by the mechanical exfoliation of bulk (natural or synthetic) crystals which leads to micrometer sized flakes, suitable for scientific investigations or for demonstrating prototype devices, but preventing their use in realistic applications produced at an industrial scale. Growth techniques, mainly chemical vapor deposition (CVD)~\cite{CVD_MoS2} and molecular beam epitaxy (MBE)~\cite{MBEMoS2,Yue2017}, have been developed to produce large scale monolayers of S-TMD suitable for electronics and optoelectronics applications, or to elaborate vertical/horizontal heterostructures~\cite{Genevieve2014,Yang2017}. In this letter, we demonstrate CVD grown S-TMD monolayers with optical quality comparable to state of the art exfoliated flakes. For that purpose, we have encapsulated it in between hexagonal boron nitride (h-BN) flakes in order to unveil its intrinsic properties. We show that the optical response of this CVD grown monolayer is comparable to the best exfoliated monolayers in the same environments, and we demonstrate this improved optical quality by performing the spectroscopy and the magneto-spectroscopy of the Rydberg states of A excitons.

\begin{figure}
\includegraphics[width=1\linewidth,angle=0,clip]{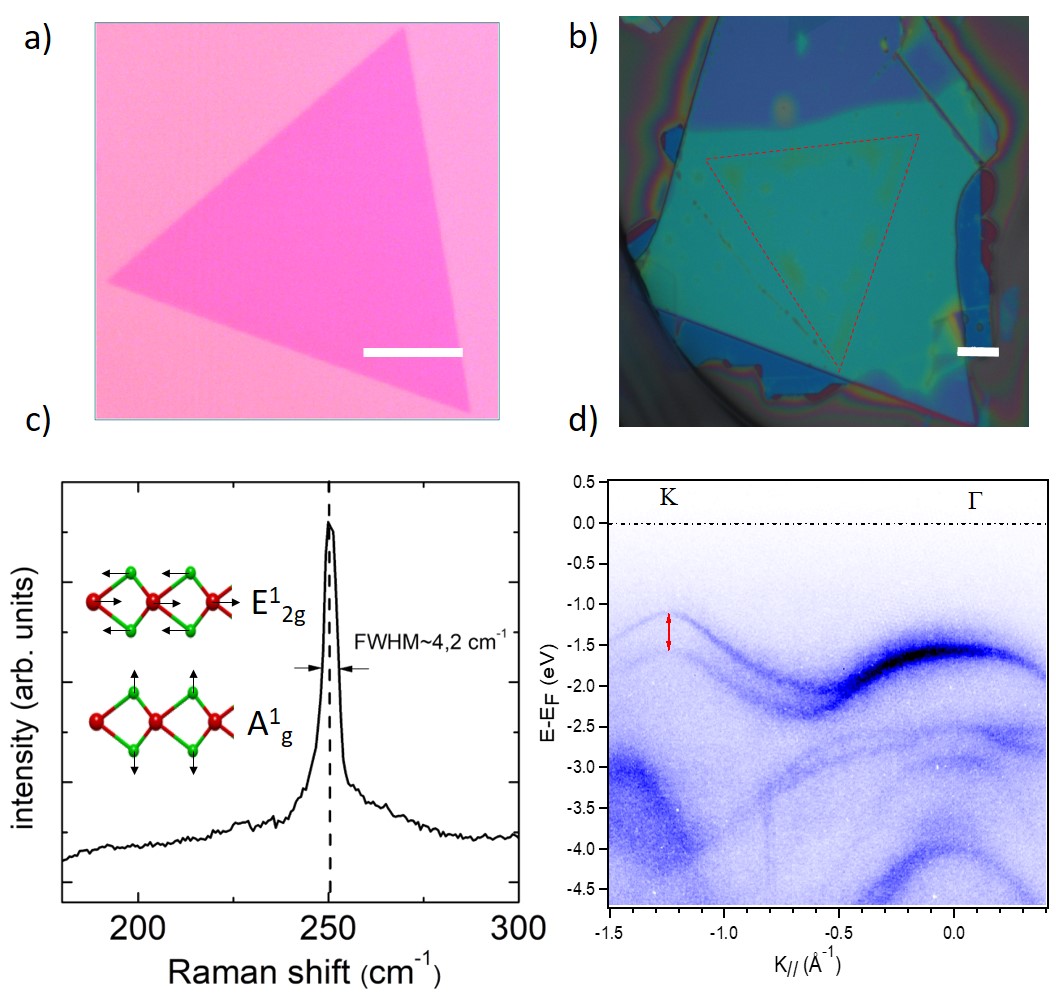}
\caption{\label{Fig1}
a) Optical image of the CVD grown monolayer on a SiO2 substrate b) Optical image of the WSe$_2$ monolayer encapsulated in h-BN. The WSe$_2$ monolayer is indicated by the dashed red lines. The scale bars in a) and b) are 10~$\mu m$. c) Raman scattering spectrum of the WSe$_2$ monolayer encapsulated in h-BN and measured with $\lambda_{exc}=515$~nm. The inset shows the atomic displacements (W atoms in red and Se atoms in green) corresponding to the $E^1_{2g}$ and to the $A_{1g}$ phonons. d) ARPES spectrum measured at $T=77$~K. The high symmetry K and $\Gamma$ points are indicated and the red double arrow indicates the $510$~meV splitting of the valence band at the K point.
}
\end{figure}

WSe$_2$ flakes were grown by chemical vapor deposition (CVD) in a quartz tube furnace on SiO$_2$. Tungsten diselenide powder was placed at the center of the furnace, and a piece of SiO$_2$/Si substrate was placed at the downstream of the quartz tube and heated to $1100$~$^{\circ}C$. Fig.~\ref{Fig1}a) and b) present two optical photographs of typical flakes produced by CVD before and after encapsulation in hBN, respectively. They extent over typically $50$ to $100$~µm. We present in Fig.~\ref{Fig1}c the room temperature Raman scattering response of the monolayer measured with an excitation laser $\lambda_{exc}=515$~nm. As already reported~\cite{Tonndorf2013}, the nearly degenerate $E^1_{2g}$ and $A_{1g}$ phonons~\cite{Luo2013} (atomic displacement pattern shown in the inset of Fig.~\ref{Fig1}c) appear around $250$~cm$^{-1}$ with a full width at half maximum (FWHM) of $4.2$~cm$^{-1}$.

We have performed band structure measurements using synchrotron radiation angle resolved photoemission spectroscopy (ARPES) at the Antares beamline (SOLEIL)\cite{Pierucci2016}. Fig.~\ref{Fig1}d displays the measured band structure around the K valley of WSe$_2$ on SiO$_2$. The top of the valence band at the K point is mostly formed by planar d$_{xy}$ and d$_{x^2-y^2}$ orbitals of tungsten, while at the $\Gamma$ point the band is mostly composed by W $d_{z^2}$ orbitals and Se $p_z$ orbitals. The observation of a single valence band at $\Gamma$ with a higher binding energy than at K also excludes the contribution from bilayer or trilayer WSe$_2$. The maximum of the valence band is located at the K point ($-1.14$~eV, which is $0.44$~eV higher than at $\Gamma$ point). The FWHM of the valence bands at the K point is of $75$~meV and the sharpness of the different bands can be attributed to the high quality of the CVD grown flake. The measured spin-orbit splitting at K is about $510$~meV, in good agreement with previous reports~\cite{Le2015}.

Optical experiments have been performed in a cold finger cryostat using a solid state diode laser at $515$~nm and a $50$~cm focal length spectrometer equipped with a nitrogen cooled charge couple device (CCD) camera. Reflectance measurements have been performed using a halogene white lamp and the contrast are calculated using the reflectance spectrum of the structure without the TMD monolayer. Magneto-optical experiments have been performed in a $20$~MW resistive magnet using an optical fiber based insert and piezo stages to move the sample below the excitation laser spot. Both experimental setups have a spatial resolution of $\sim 1\mu m$. Polarization of the photoluminescence (PL) signals was analyzed using a quarter-wave plate and a linear polarizer. PL experiments essentially probe the low energy A exciton but it has been shown that in monolayers of S-TMD, exciton excited states several hundreds of meV above the exciton ground state can also give rise to PL emission~\cite{manca2017}. Higher energy excitons such as the B exciton are investigated with reflectance spectroscopy or with non-linear optical techniques~\cite{wang2015,Han2018}.

\begin{figure}
\includegraphics[width=1\linewidth,angle=0,clip]{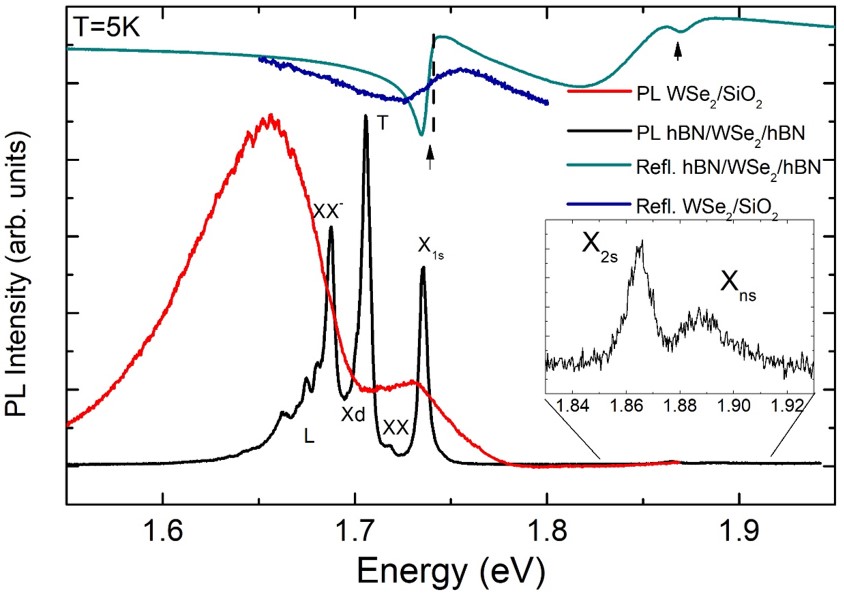}
\caption{\label{Fig1bis}
PL spectrum measured at $T=5K$ of the WSe2 flake on a SiO$_2$ substrate (red solid line) and after encapsulation in h-BN (black solid line). Some excitonic complexes are indentified: the biexciton XX, the trion T, the grey exciton (X$_d$), and the charged biexciton (XX-). The green curve is a reflectance contrast spectrum of the encapsulated flake and the two arrows point at the energy positions of the $1s$ and of the $2s$ states signatures. The blue curve is the reflectance spectrum of the WSe$_{2}$ monolayer on SiO$_2$. The inset is a zoom on the high energy range where the $2s$ Rydberg state is observed.
}
\end{figure}

The photoluminescence (PL) and reflectance spectra of the as-grown WSe$_2$ monolayer on SiO$_2$ are presented as the red and blue curves in Fig.~\ref{Fig1bis}, respectively. The PL shows a 40 meV wide A exciton emission at 1.717 meV and a broad band, attributed to exciton complexes (biexciton, trion, etc ..) and to localized excitonic states, at lower energy. One can note a pronounced Stoke shift of 15 meV, of the PL peak with respect to the resonance measured in reflectivity. As will be shown in the following, the observed linewidth of the exciton feature is highly misleading and does not reflect the intrinsic quality of the material but is more representative of its interaction with the SiO2 substrate. These are the only observable features in the PL spectrum. Excitons are electron-hole pairs bound by Coulomb interaction. They have an internal energetic structure composed of a ground state indexed by a principal quantum number $n=1$ and a series of excited states $n=2,3,4, ...$. The sequence of excited states is defined by the potential acting on the electron-hole pair. In monolayers of S-TMD, it was shown that the potential is a screened Coulomb potential, the Keldysh potential~\cite{Keldysh1979}, due to the reduced dimensionality and of the high polarizability of the S-TMD monolayer with respect to the dielectric environment~\cite{He2014,chernikov2015}. These excited states can be observed in the optical response of high quality and neutral monolayers of S-TMDs, i.e. in their reflectance and photoluminescence spectra. To improve the optical quality of the CVD grown WSe$_2$ monolayer, we have encapsulated it with two thin layers of h-BN using a dry transfer technique~\cite{gomez}. A polypropylene carbonate(PPC) stamp was prepared on a glass slide and was used to pick-up first the top h-BN layer, then the CVD grown WSe$_2$ and finally this stack was deposited on the bottom h-BN layer on a $300$~nm thick SiO$_2$/Si substrate. As already reported for exfoliated monolayers of S-TMD~\cite{CadizPRX}, we observe profound changes in the emission spectrum of our CVD grown monolayer after encapsulation in h-BN.

These changes are evidenced in Fig.~\ref{Fig1bis} which shows the PL spectrum after encapsulation (black curve): most noticeable is the linewidth of A exciton $1s$ peak which is strongly reduced down to $4.7$~meV, the resonance corresponding to the exciton $1s$ peak in reflectance is red shifted by $4$~meV as a result of the change in the dielectric properties of the surrounding environment (SiO$_2$ and vaccuum with respect to h-BN), a new structure is observed at higher energies close to $1.87$ eV which we attribute to the $2s$ excited state of the A exciton~\cite{stier2018,molas2019}. Finally, the broad band at lower energy acquires a rich structure of discrete peaks associated to~\cite{ye2018} the biexciton XX, the singlet and triplet trions T, possibly the dark exciton, the charged biexciton XX-, and finally to localized excitons L. In Fig.~\ref{Fig1bis}, we present also a reflectance contrast spectrum of the encapsulated WSe$_2$ (green curve). The $1s$ and $2s$ resonances are indicated by black arrows and match nicely with the observed peaks in the PL spectrum, confirming our initial assignment. The energy difference between the $1s$ and $2s$ states is of $130$~meV, in line with values reported for WSe$_2$ in a similar dielectric environment~\cite{stier2018,molas2019}. Exciton excited states are only visible in the emission spectrum of the encapsulated monolayer, and are absent from the spectrum of the monolayer on SiO$_2$ as a result of the improved optical quality and of the charge neutrality provided by the encapsulation in the clean and inert h-BN layers. Exciton excited states with higher indices are not directly observable in the spectrum, possibly because their energy spacing is smaller than their linewidth. They only appear in the form of a broad feature at energies above the $2s$ emission and labelled $ns$ in Fig.~\ref{Fig1bis}. A magnetic field should separate them~\cite{stier2018,molas2019} and allow for their observation, but, in the present case, their emission is too weak to be detected using our optical fiber based magneto-optical spectroscopy set-up.

\begin{figure}
\includegraphics[width=0.7\linewidth,angle=0,clip]{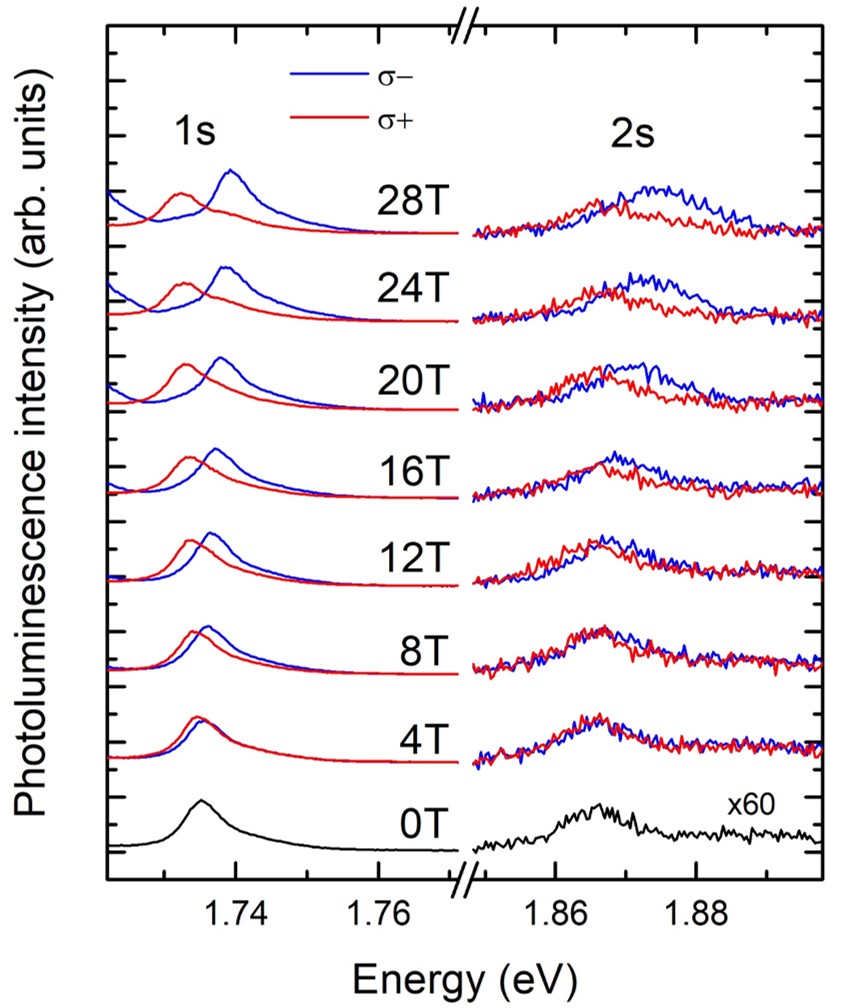}
\caption{\label{Fig2}
Polarization resolved photoluminescence spectra at selected values of the magnetic field (red curves for $\sigma^+$ and blue curves for $\sigma^-$ polarizations) for the 1s and 2s (multiplied by 60) states.
}
\end{figure}

To confirm the origin of these different peaks in the PL spectrum of our encapsulated CVD grown WSe$_2$ monolayer, we have performed polarization resolved magneto-PL experiments in the Faraday configuration. When a magnetic field is applied perpendicular to the plane of the WSe$_2$ monolayer, K$^+$ and K$^-$ valleys show an opposite energy shift, the valley Zeeman effect\cite{Aivazian2014, Srivastava2014, MacNeill2015}. This effect is observed in our data as a splitting of the exciton $1s$ peak in opposite polarizations. This effect, presented in Fig.~\ref{Fig2}, allows for the determination of a fundamental parameter characterizing the exciton, the excitonic Landé factor~\cite{Aivazian2014,stier2018,molas2019} $g=-4.3$ for the $1s$ resonance, and is slightly reduced to $g=-4.1$~for the $2s$ resonance.

\begin{figure}
\includegraphics[width=0.7\linewidth,angle=0,clip]{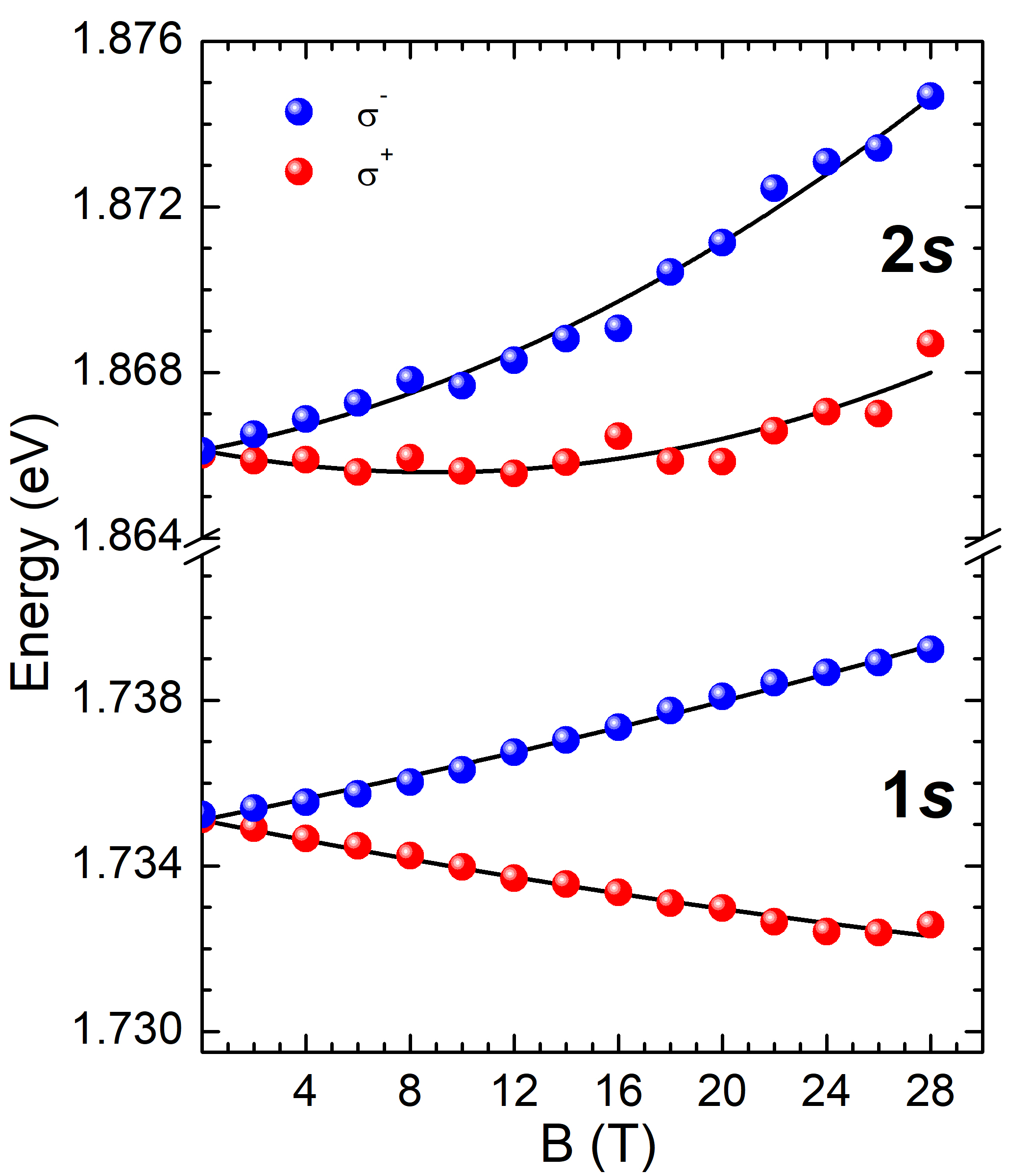}
\caption{\label{Fig3}
Central energy position of A exciton 1s and 2s states as a function of the magnetic field fior the two $\sigma^{\pm}$ polarizations. The solid black line are the results of fits including both the Valley zeeman effect and the diamagnetic shift.
}
\end{figure}

Finally, different excitonic states are characterized by different spatial extents which will result in distinct quadratic diamagnetic shifts \mbox{$\Delta E_{dia}=\frac{e^2}{8m_r}<r^2>B^2=\sigma B^2$}, where $m_r$ is the exciton reduced mass ($m_r=0.2 m_0$~\cite{stier2018}), and $r=\sqrt{8 m_r \sigma}/e$ is the radial coordinate perpendicular to B. Fig.~\ref{Fig3} shows the energy position of the polarization resolved $1s$ and $2s$ emission peaks as a function of the magnetic field. While the $1s$ peak mainly shows a splitting with a magnitude linear in B related to the valley zeeman effect, a quadratic B dependence is clearly observed for the $2s$ state. This indicates a much reduced spatial extent for the wave function of the $1s$ state than the one of the $2s$ state. From these measurements, one can extract $\sigma_{1s} \sim 9\times10^{-7} (\pm 8\times 10^{-8})$~$eV.T^{-2}$ and $\sigma_{2s} \sim 6.6\times 10^{-6}(\pm 2\times 10^{-7})$~$eV.T^{-2}$, corresponding to $r_{1s}\sim 2.7\pm 0.2$~nm and $r_{2s}\sim 7.7\pm 0.1$~nm. The radius corresponding to the $1s$ state deduced from our experiment is significantly larger than the one obtained with a similar sample at higher magnetic fields~\cite{stier2018}.  We speculate that this discrepancy could be related to the large thickness of the h-BN encapsulating layers, 160 nm and 25 nm for our top and bottom hBN layers, respectively, changing the average dielectric constant of the environment of the flake. Additionally, one should consider possible variations of the exciton reduced mass, the value of which also depends on the dielectric environment. The radius corresponding to the $2s$ state can be deduced with more accuracy and it matches values reported so far for $2s$ states in WSe$_2$ encapsulated in h-BN. These values are in clear disagreement with expectations for the 2D hydrogen model~\cite{MacDonald1986} and reflect the peculiar electrostatic screening in monolayers of S-TMD.

To conclude, we have demonstrated CVD grown WSe$_2$ with optical quality reaching the one of exfoliated samples. This high optical quality is evidenced by a linewidth as small as $4.7$~meV once encapsulated in h-BN for the emission from the $1s$ state of A excitons and by the observation of emissions from the $2s$ state and from higher index states. Magneto-optical spectroscopy confirms the origin of these emissions by allowing the measurement of their g-factor and of their diamagnetic shift. Our work shows that CVD grown monolayers of S-TMD are promising candidates for optoelectronic applications requiring large scale growth of materials.

\begin{acknowledgements}
This work was partially supported by the EU Graphene Flagship Project (Project No. 785219), by the French National Research Agency (ANR) in the framework of the J2D project (ANR-15-CE24-0017), grants Labex Nanosaclay (ANR-10-LABX-0035) and RhomboG (ANR-17-CE24-0030), and in the framework of the Investissements d’Avenir program (ANR-15-IDEX-02), and by the National Natural Science Foundation of China No. 51525202. Growth of hexagonal boron nitride crystals was supported by the Elemental Strategy Initiative conducted by the MEXT, Japan and the CREST (JPMJCR15F3), JST. We thank the Nanofab group at Institut Néel for help with van der Waals heterostructures preparation setup. We thank Dr. Jose Avila and Dr. Maria C. Asencio for the ARPES experiments.
\end{acknowledgements}

\end{document}